\documentclass[aps,showpacs,superscriptaddress,twocolumn]{revtex4}

\usepackage{graphicx}

\begin{document}
                       
\title{Magnetic susceptibility in quasi one-dimensional \mbox{\rm
Ba$_2$V$_3$O$_9$}: chain segmentation versus the staggered field
effect}

\author{B.~Schmidt}
\email{bs@cpfs.mpg.de}
\affiliation{Max Planck Institute for the Chemical Physics of Solids, 
Dresden, Germany}
\author{V.~Yushankhai}
\affiliation{Max Planck Institute for the Chemical Physics of Solids, 
Dresden, Germany}
\affiliation{Joint Institute for Nuclear Research, Dubna, Russia}
\author{L.~Siurakshina}
\affiliation{Max Planck Institute for the Physics of Complex Systems,
 Dresden, Germany}
\affiliation{Joint Institute for Nuclear Research, Dubna, Russia}
\author{P.~Thalmeier}
\affiliation{Max Planck Institute for the Chemical Physics of Solids, 
Dresden, Germany}

\pacs{75.10.Jm, 75.40.Cx, 75.50.Ee}

\bibliographystyle{apsrev}

\begin{abstract}
    A pronounced Curie-like upturn of the magnetic susceptibility
    $\chi(T)$ of the quasi one-dimensional spin chain compound
    Ba$_2$V$_3$O$_9$ has been found recently~\cite{kaul:02}. 
    Frequently this is taken as a signature for a staggered field
    mechanism due to the presence of g-factor anisotropy and
    Dzyaloshinskii-Moriya interaction.  We calculate this contribution
    within a realistic structure of vanadium $3d$- and oxygen
    $2p$-orbitals and conclude that this mechanism is far too small to
    explain experimental results.  We propose that the Curie term is
    rather due to a segmentation of spin chains caused by broken
    magnetic bonds which leads to uncompensated S=$\frac{1}{2}$ spins
    of segments with odd numbers of spins.  Using the finite-temperature
    Lanczos method we calculate their effective moment and show that
    $\sim1\,\%$ of broken magnetic bonds is sufficient to reproduce
    the anomalous low-$T$ behavior of $ \chi(T)$ in Ba$_2$V$_3$O$_9$.
\end{abstract}

\maketitle
  
\section{Introduction}
Recently a quasi one-dimensional magnetic behavior of Ba$_2$V$_3$O$_9$
was clearly revealed by means of the magnetic susceptibility $\chi(T)$
and the specific heat $C_{\rm p}(T)$ measurements in polycrystalline
samples~\cite{kaul:02}.  The data are compatible with the spin $S=1/2$
antiferromagnetic (AF) Heisenberg chain model with a nearest neighbor
exchange coupling $J=94\,\mbox{K}$.  In addition an anomalous
Curie-like upturn of $\chi(T)$ was found below 20\,K and claimed to be
of intrinsic nature because the effect of paramagnetic impurities was
ruled out by the analysis of experimental data.  This low-$T$ behavior
of $ \chi(T) $ was tentatively attributed in~\cite{kaul:02} to the
staggered field effect induced by the applied magnetic field.  In a
quasi one-dimensional spin $S=1/2$ chain a low-$T$ upturn of $\chi(T)$
is expected if the Dzyaloshinskii-Moriya (DM) interaction and/or a
staggered $g$-factor anisotropy are present in the
system~\cite{oshikawa:97,affleck:99}.  Among the $3d$-systems, Cu
benzoate~\cite{dender:96} and pyrimidine Cu
dinitrate~\cite{feyerherm:00} are the most known examples.  However
the anomalous low-$T$ part of $\chi(T)$ in Cu benzoate is much larger
than in the theory~\cite{affleck:99} and the physical reason for this
discrepancy is not yet clear.  For better understanding, a comparative
analysis of different sources contributing to the anomalous low-$T$
behavior of $\chi(T)$ is necessary.  In the present paper such an
analysis is developed for Ba$_2$V$_3$O$_9$ compound.

The low symmetry of Ba$_2$V$_3$O$_9$ allows both a staggered
$g$-factor anisotropy and a DM interaction to contribute to the
staggered field.  We suggest a simple model of vanadium $3d$- and
oxygen $2p$-orbital structure within magnetic chains and calculate
both contributions.  Alternatively, we consider a segmented spin chain
model where uncompensated spin moments lead to the low- $T$ Curie term
and compare the results of both models to the experimental $\chi(T)$. 
A segmented AF spin-1/2 chain model was applied
earlier~\cite{asakawa:98,kiryukhin:01} to explain the magnetic
susceptibility measurements in quasi one-dimensional cuprates.  In
these studies, analytic results~\cite{asakawa:98} and quantum Monte
Carlo simulations~\cite{kiryukhin:01} were used, assuming a particular
random distribution of broken magnetic bonds (defects).  Here we
extend the segmented-chain model and compare different distributions 
of defects or chain segment lengths. To determine the thermodynamic 
properties of the chains under consideration, we apply the 
finite-temperature Lanczos method~\cite{jaklic:00}.

\section{Orbital structure and the $g$-tensor anisotropy}

\begin{figure}
    \includegraphics[width=\columnwidth]{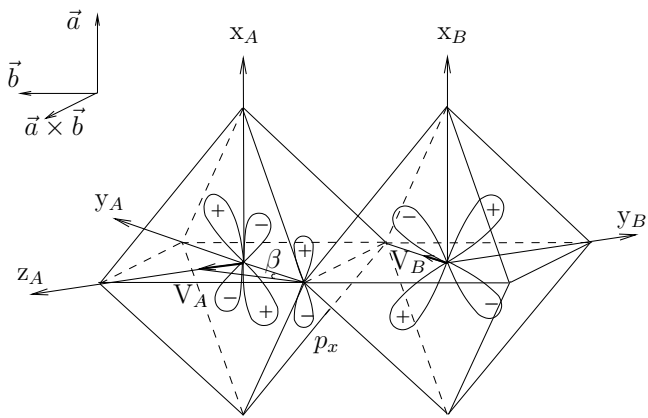}
    \caption{Two edge-sharing VO$_6$ octahedra forming a segment of a
    magnetic chain in Ba$_2$V$_3$O$_9$.  The local coordinate systems
    attached to the A- and B-octahedra and the directions of the
    off-center displacements of the V$_{\rm A,B}$-ions along the local
    $z$-axes are shown.  The ground-state $d_{xy}^{\rm A}$ and
    $d_{xy}^{\rm B}$ orbitals of the V$_{\rm A,B}$-ions and the
    relevant $p_x$-orbital of a common-edge oxygen ion responsible for
    the V$_{\rm A}$--V$_{\rm B}$ superexchange are shown.
    The $d_{xy}$ orbitals in neighboring octahedra lie in orthogonal
    planes. 
    The Dzyaloshinskii-Moriya vector ${\bf D}_{\rm AB}$ is parallel to
    ${\bf a}$-axis and alternates in sign on successive bonds.
    }
    \label{fig:fig1}
\end{figure}
According to~\cite{kaul:02} VO$_6$ octahedra in Ba$_2$V$_3$O$_9$ form
edge-sharing chains. The O$_6$ octahedron with an average V--O
distance $\simeq2\,\mbox{\AA}$ is
only slightly distorted, and V$^{4+}$ ion ($d^1$-state) is displaced
from the center by about $0.2\,\mbox{\AA}$ towards one of the oxygens
shared by two neighboring octahedra.  The direction of V-ion
off-center displacement distinguishes clearly a local $z$-axis in each
VO$_6$ octahedron.  The energy splitting of vanadium $d$-orbitals was
suggested~\cite{kaul:02} to be such that $d_{xy}$-orbital located in
the plane transverse to the local $z$-axis is the singly occupied
ground state orbital (Figure~\ref{fig:fig1}). Results of {\em ab initio}
calculations of a small cluster confirm this picture. 
Within a chain the local $z$-axis varies in a zig-zag manner thus
forming alternating short and long V--O bonds.  The
resulting $d$-orbital arrangement along a chain determines the
symmetry and strength of a superexchange (SE) coupling between
vanadium spins.  Based on standard SE theory we
calculate both the isotropic exchange constant $J_{ij}$ and the DM
vector ${\bf D}_{ij}$ for neighboring V-ions.

First we calculate the $g$ tensor components of the V$^{4+}$ ion in
the VO$_{6}$ octahedron whose size is scaled to the one in
Ba$_2$V$_3$O$_9$.  Strong variation of V--O covalent bonding due to a
V-ion displacement from the central position in the nearly cubic O$_6$
cage splits $t_{2g}$ and $e_g$ vanadium $d$-orbitals.  We neglect the
extra distortion of the crystal field on V-ions caused by the side
ionic groups attached to the chains.  Then the diagonal $g$ tensor is
given by $ g_{\parallel} = 2(1- \Lambda_{zz}), g_{\bot}=2(1-
\Lambda_{\bot})$, where $\Lambda_{\bot}=\Lambda_{xx}=\Lambda_{yy}$,
and $\Lambda_{\mu\mu}=\lambda\sum_{m\neq 0}|\langle 0| L_\mu |m\rangle
|^2/( E_m-E_0)$.  Here, $\lambda \simeq 0.03\,\mbox{eV}$ is the
constant of spin-orbit coupling for V$^{4+}$-ions~\cite{abragam:70},
$L_{\mu}$ is the orbital angular momentum and $| 0\rangle$, $| m
\rangle$ are ground state and an excited d-orbital state respectively
with an excitation energy $E_m-E_0\gg\lambda$.

\begin{figure}
    \includegraphics[width=.8\columnwidth]{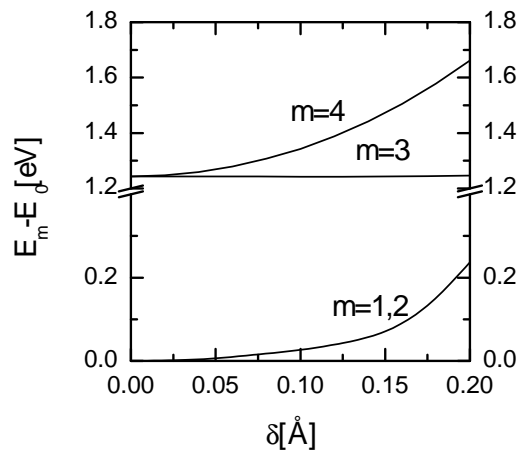}
    \caption{Excitation energies $E_{m}-E_{0}$ of the Vanadium $3d$ 
    states $|m\rangle$ in VO$_{6}$ as functions of the off-center 
    displacement $\delta$ of the V$^{4+}$-ion along the local 
    octahedral $z$-axis.}
    \label{fig:fig3}
\end{figure}
For a VO$_6$ octahedron with one unpaired electron on vanadium d-shell
we performed {\em ab initio\/} quantum chemical calculations of the
ground state $E_0$ and excited state $E_m$ cluster energies as
function of the V$^{4+}$ ion displacement $\delta $ along a local
$z$-axis.  The model cluster [VO$_6$]$^{8-}$ was embedded in an
octahedral point charge environment and the program package {\sc
molpro} was used.  First the restricted Hartree-Fock molecular orbital
(MO) wave functions were determined to define an active space of MOs
including a variety of valence and virtual orbitals (altogether 23
orbitals arising from strongly hybridized V(3$d$) and O(2$p$) states). 
Then the active space was used to take into account effects of
electron correlations by performing multireference configurational
interaction ({\sc MRCI}) calculations of the cluster energies when the
unpaired electron was placed successively in different $d$-orbitals. 
The results of these post Hartree-Fock calculations
(Figure~\ref{fig:fig3}) show that $t_{2g}$-orbitals are split into the
ground state orbital singlet $| d_{xy}\rangle\equiv | 0\rangle$ and
degenerate orbital doublet $| d_{xz}\rangle\equiv | 1\rangle$, $|
d_{yz}\rangle\equiv | 2\rangle$.  At the physical displacement $\delta
= $0.2 \AA, the energy difference between them is $E_{1,2} - E_0
\simeq 0.23\,\mbox{eV}\equiv\Delta_1$, while for $| d_{x^2-y^2} \equiv
| 3\rangle$ one has $E_3 - E_0 \simeq 1.26\,\mbox{eV}\equiv \Delta_3$. 
Using $\langle 1| L_x| 0\rangle = - \langle 2 | L_y | 0\rangle = i$
and $\langle 3 | L_z | 0\rangle = -2i$ one obtains the following
result for the $g$-factor anisotropy: $(g_{\parallel} - g_{\bot})
\simeq 2\lambda (1/\Delta_1 - 4/\Delta_3) \simeq 0.07$.

Next we define a global reference system with unit vectors ${\bf\hat
x} \parallel {\bf a}$, ${\bf\hat y} \parallel {\bf b}$ and ${\bf\hat
z} \parallel {\bf a} \times {\bf b}$ , where $\bf a$ and $\bf b$ are
orthogonal crystallographic axes in Ba$_2$V$_3$O$_9$.  In this system
the $\hat g$ tensor is decomposed into a uniform $\hat g_{\rm u}$ and
a staggered $\hat g_{\rm s}$ part, ${\hat g}_{\ell}= \hat g_{\rm u} +
(-1)^\ell\hat g_{\rm s}$, where $\ell$ denotes the V-sites along a
chain.  The only non-zero components of $\hat g_{\rm s}$ are $(\hat
g_{\rm s})_{yz}=(\hat g_{\rm s})_{zy}=g_{\rm s}= (g_{\parallel} -
g_{\bot})\sin{\alpha}\cos{\alpha} $, where $\alpha$ is a tilt angle. 
In Ba$_2$V$_3$O$_9$ the local $z$-axes are tilted from the chain
direction ($\bf b$-axis) and $\alpha \simeq 45^{\circ}$ alternates
as $\ell$ runs along the chain.  Finally, our calculation leads to a
staggered component $g_{\rm s}\simeq0.035$ for Ba$_2$V$_3$O$_9$ .

\section{Isotropic exchange and Dzyaloshinskii-Moriya interaction}

Considering now the intrachain magnetic coupling we note that for each
(AB)-pair of neighboring V-ions the shortest SE path is via the
out-of-plane $p_x$-orbital which belongs to one bridging oxygen as shown
in Figure~\ref{fig:fig1}.  To estimate the electronic hopping
parameter $t_{d \pi}$ of the $\pi$-bonding between each of occupied
$d_{xy}^{A,B}$-orbitals and bridging oxygen $p_x$-orbital we use
Harrison's prescription~\cite{harrison:89}: $t_{d \pi}=\eta_{pd \pi}
\hbar^2r_d^{3/2}/(ma^{7/2})$, where $\eta_{pd \pi}=1.36$,
$\hbar^2/m=7.62\,\mbox{eV\AA}$ and $r_d=0.98\,\mbox{\AA}$ for
vanadium.  For the average V--O distance $a \simeq 2\,\mbox{\AA}$
one obtains $t_{d \pi}\simeq1\,\mbox{eV}$. Then from SE theory the
isotropic exchange constant is obtained as
\begin{equation}
    J\simeq\frac{4t^4_{d\pi}}{\left(U_d+\Delta_{d\pi}\right)^2}
    \left[\frac{1}{U_d}+\frac{2}{2(U_d+\Delta_{d\pi})+U_{\pi}}\right],
\end{equation}
where $U_d$ and $U_{\pi}$ are the on-site Coulomb integrals on V and
O, respectively; $\Delta_{d\pi}=\epsilon_d - \epsilon _{\pi}$ is the
energy difference between the vanadium ground state $d$- and oxygen
$p_x$-orbitals.  We have $\Delta _{d \pi} >0$, since Ba$_2$V$_3$O$_9$
is a Mott-Hubbard insulator.  We use the following parameter values
(in eV): $U_d=3\ldots4$, $U_{\pi}=2\ldots6$, and $\Delta_{d \pi}
\simeq 5$, characteristic of vanadium oxides with tetravalent V-ions. 
This leads to an exchange $J$ in the range $10\,\mbox{meV}<
J<20\,\mbox{meV}$ in comparison to the experimental value
$J\approx10\,\mbox{meV}$.  We conclude that the use of the shortest SE
path via the common edge O-ion provides a satisfactory account of the
isotropic AF exchange in Ba$_2$V$_3$O$_9$.

For a pair (AB) of V-ions the DM vector can be written
\cite{keffer:62} as ${\bf D}_{\rm AB} = -i \left({\bf\Lambda}_{\rm A}
- {\bf\Lambda}_{\rm B}\right)$ with $ {\bf\Lambda}_{\rm A/B} =
2\lambda \sum_{m\neq 0}\langle m_{\rm A/B}| {\bf L}_{\rm A/B}|0_{\rm
A/B}\rangle(E_m-E_0)^{-1} J_{\rm A/B}^{(m)}$.  Here, $J_{\rm
A}^{(m)}=J(m_{\rm A}, 0_{\rm B}; 0_{\rm A}, 0_{\rm B})$ and $J_{\rm
B}^{(m)}=J(0_{\rm A}, m_{\rm B}; 0_{\rm A}, 0_{\rm B} )$ are SE
constants in similar intermediate configurations where the unpaired
electron on the V$_{\rm A}$- or V$_{\rm B}$-ion, respectively, is
raised by the spin-orbit interaction to $m$th excited $d$-state.  The
symmetry requires $J(m_{\rm A}, 0_{\rm B}; 0_{\rm A}, 0_{\rm B} ) =
J(0_{\rm A}, m_{\rm B}; 0_{\rm A}, 0_{\rm B} ) \equiv J_{\rm
AB}^{(m)}$.  In this notation, the isotropic constant calculated above
reads $ J = 2J(0_{\rm A}, 0_{\rm B}; 0_{\rm A}, 0_{\rm B} ) \equiv
2J_{\rm AB}^{(0)}$.  For the orbital geometry in
Figure~\ref{fig:fig1}, inspection shows that the only non-zero excited
state exchange $J_{\rm AB}^{(m)}$ is for $m=1$.  This corresponds to
the intermediate electronic configuration with occupied $| d_{xz}
\rangle$ state.  The matrix elements of $ {\bf L}_{\rm A}$ and $ {\bf
L}_{\rm B} $ are calculated by referring to a common coordinate
system, which yields $\mbox{$\langle1_{\rm A}|{\bf L}_{\rm A}|0_{\rm
A}\rangle$}-\mbox{$\langle1_{\rm B}|{\bf L}_{\rm B}|0_{\rm
B}\rangle$}=-2i{\bf\hat e}_{\rm AB}$, where the unit vector ${\bf\hat
e}_{\rm AB}$ is defined as ${\bf\hat e}_{\rm AB} = {\bf\hat z}_{\rm A}
\times {\bf\hat z}_{\rm B}$.  To compare $J_{\rm AB}^{(1)}$ and
$J_{\rm AB}^{(0)}$, we note that the hopping parameter $t_{d
\pi}^{(1)}$ between each of the excited vanadium $d_{xz}^{\rm
A,B}$-orbitals and the bridging oxygen $p_x$-orbital is related to
$t_{d \pi}$-hopping as $t_{d \pi}^{(1)} \simeq t_{d \pi} \sin \beta$,
where $t_{d \pi}$ enters into the definition of $J_{\rm AB}^{(0)}$ and
$\sin \beta \simeq (\delta/a) \simeq 0.1$.  This immediately leads us
to the simple result, $J_{\rm AB}^{(1)} \approx J_{\rm AB}^{(0)} \sin
\beta$, hence ${\bf D}_{\rm AB}/J_{\rm AB} = d \cdot {\bf\hat e}_{\rm
AB}$, where $d=(2\lambda / \Delta_1) \sin \beta \approx 0.025$.
 
${\bf D}_{\rm AB}$ is staggered along the chain
direction, i.e. ${\bf D}_{\ell,\ell+1}=(-1)^\ell{\bf D}$ due to the
property of ${\bf e}_{\ell,\ell+1}$. The geometric factor
$\sin\beta$ is a measure of the asymmetry caused by off-center
displacements of V-ions. If they are neglected
the inversion center at the midpoint of (AB)-pair is
restored and ${\bf D}_{\rm AB}=0$.

\section{Staggered-field model for the susceptibility upturn in
Ba$_2$V$_3$O$_9$}

Taking both effects described above into account, a magnetic field
$\bf H$ applied to a chain induces a staggered field $\bf h$, which
can approximately be written~\cite{oshikawa:97,affleck:99} as
\begin{equation}
    {\bf h}\approx\frac{1}{J}{\bf D}\times{\bf H} +
    \hat g_{\rm s}{\bf H}.
\end{equation}
For $T\ll J/k_{\rm B}$, the staggered field leads to a contribution to
the magnetic susceptibility of the form~\cite{affleck:99}:
\begin{equation}
    \chi_{\rm s}(T)\simeq0.2779c^2\left(
    \frac{N_{\rm A}\mu_0\mu_{\rm B}^2}{k_{\rm B}}\right)
    \frac{\ln^{1/2}(J/k_{\rm B}T)}{T},
\end{equation}
which scales with the factor $c\sim h/H$\cite{oshikawa:97}.  This should
be compared with the experimental low-$T$ ($2\,\mbox K<T<20\,\mbox
K$) behavior of $\chi$ in Ba$_2$V$_3$O$_9$ which is well described by a
Curie law, $\chi_{\rm LT}=C_{\rm LT}/T$\cite{kaul:02}.

The staggered susceptibility $\chi_{\rm s}(T)$ varies approximately
like $C_{\rm s}/T$.  For $2\,\mbox K<T<20\,\mbox K$, we take into
account the slowly varying logarithmic correction by replacing
$\ln^{1/2}(J/k_{\rm B}T)$ with its average value $\approx1.6$.  Then,
we obtain the Curie constant due to the staggered field as $C_{\rm
s}\simeq1.33c^2(N_{\rm A}\mu_0\mu_{\rm B}^2/3k_{\rm B}) = (N_{\rm
A}\mu_0/3k_{\rm B})(\mu^{\rm eff}_{\rm s})^2$, where the second relation
defines the effective staggered magnetic moment $\mu^{\rm eff}_{\rm
s}=\sqrt{1.33}c \mu_{\rm B}$.

In general the factor $c^2$ is an angle dependent function of the
magnetic field $\bf H$ direction:
$c^2(\theta,\phi)=\sum_{\mu}\left(\partial h_{\mu}/\partial
H_{\alpha}\right)^2$, $\alpha=\alpha(\theta,\phi)$.  In a polycrystal
used in~\cite{kaul:02}, the angular average $c^2= \left\langle
c^2(\theta, \phi)\right\rangle$ is measured.  With $\bf D$ and $\hat
g_{\rm s}$ given, one obtains $c^2=(2/3)[d^2 + g_{\rm s}^2]$, yielding
$C_{\rm s} \simeq 2.5\cdot 10^{-3}\,\mbox{cm$^3$K/mol}$, and
correspondingly an effective moment $\mu^{\rm eff}_{\rm s} \simeq
4\cdot 10^{-2} \mu_{\rm B}$.

In comparison the experimental values reported in~\cite{kaul:02},
namely $C_{\rm LT}\simeq6.0\cdot10^{-2}\,\mbox{cm$^3$K/mol}$ and
$\mu^{\rm eff}_{\rm LT}\simeq0.2\mu_{\rm B}$, are much higher. 
Therefore we conclude that the staggered-field effect alone is not
sufficient to explain the low-temperature behavior of the magnetic
susceptibility in Ba$_2$V$_3$O$_9$, and a different mechanism must be
present.

\section{Segmented-chain model for the anomalous susceptibility}

The previous magnetic model implies a long-range order of short V--O
bonds in each structural VO$_2$-chain.  From Figure~\ref{fig:fig1} we
infer that several quasi-degenerate energy minima for the short V--O
bonds orientation can exist.  
This expected structural degeneracy is described in an easy way by
recalling that a local $z$-axis in each VO$_6$ octahedron is defined
in our description by the direction of the off-center V-ion
displacement as is shown in Fig.~\ref{fig:fig1}.  Then, for a separate
chain, four degenerate ground state structural configurations
correspond to the following decomposition of the local $z$-axes and
their variation along a chain: ${\bf\hat z}_{\ell}= \pm [{\bf\hat
b}\cos{\alpha} \pm (-1)^\ell ({\bf\hat a} \times {\bf\hat b})
\sin{\alpha} ]$, with the tilt angle $\alpha \simeq 45^{\circ}$. 
Because of weak interchain interactions and the influence of the side
ionic groups, the exact degeneracy can be partly removed.  We
suggest, however, that the remaining degeneracy can lead
to a domain wall formation withing a chain, which does not contradict
the X-ray measurements reported in \cite{kaul:02}.
Chain segments with differently oriented bonds (domains) can
spontaneously form at high temperatures during the growth and
preparation of the sample.
A chain domain wall with parallel neighboring local $z_A$- and $z_B$-
axes corresponds to the ground-state $d_{xy}^{\rm A}$ and 
$d_{xy}^{\rm B}$ orbitals that lie in parallel planes and thus have no 
short-parth  SE  connection. Therefore, 
the exchange interactions between end point spins of neighboring
segments are strongly suppressed resulting in broken magnetic bonds
which we call defects below.

Chain segmentation is an attractive possibility to explain the
observed low-temperature Curie term in the magnetic susceptibility
because segments with an odd number of spins lead to uncompensated
$S=1/2$ states.  In the following we calculate this contribution.  The
effective Hamiltonian used is the one-dimensional isotropic AF
Heisenberg model, $H = J\sum_\ell{\bf S}_\ell{\bf S}_{\ell+1}$, and we
ignore the staggered-field terms discussed before in the following
analysis.

For the joint probability of having no defect between positions $0$
and $x$, but one defect at position $x$ on a particular chain, or
the probability to find a spin chain of length $x$, we use a
generalized Poisson distribution of the form
\begin{equation}
    P(x) = f(x)\exp\left(-\int_0^xf(t)dt\right).
\end{equation}
Here, $f(x)$ describes the differential probability to find a defect
at the distance $x$ measured from a defect located at the origin.  For
the standard Poisson distribution, $f(x)=1/\rho$ is just a constant,
i.\,e., is independent from the distance to the origin.

A Poissonian distribution of chain lengths overemphasizes the number
of short chains.  Instead, we expect that their fraction is small and
the distribution peaks at a finite value because defects which are
close by ``repel'' each other due to their larger elastic energies. 
Lacking a detailed microscopic model for this mechanism, we simply
assume that the probability of defects at a distance $x$ grows
linearly with $x$ like $f(x)=\frac{\pi}{2}\rho^2x$, leading to
\begin{equation}
    P(x) = \frac{\pi}{2}\rho^2x\exp\left(-\frac{\pi}{4}\rho^2x^2\right).
\end{equation}
Here, $\rho$ is the number of broken bonds per unit length of an
infinite chain, i.\,e., the inverse average chain length.  This is the
Wigner distribution, well-known from quantum statistics.

\section{Numerical procedure and results}

We use the finite-temperature Lanczos method~\cite{jaklic:00} to
calculate the eigenvalues, eigenvectors, and thermodynamic properties
for the segmented chains.  To comply with the effect of broken bonds,
we use open boundary conditions.  Then the only symmetries
remaining to reduce the size of the problem are $\left[H,{\bf
S}^2\right]=0$ and $\left[H,S_z\right]=0$.  While the latter can be
implemented easily using an appropriate basis, there is no
simple way to efficiently incorporate the former. To compute
finite-temperature expectation values, we have to store {\em all\/}
calculated Lanczos eigenvalues and {\em all\/} of the corresponding
eigenvector coefficients in terms of the basis, which is the limiting
computational requirement here, in contrast to ground-state calculations.

We have calculated the uniform magnetic susceptibility $\chi_N$ for
Heisenberg chains with length $N$ between two and 24 sites, defined by
\begin{equation}
    \chi_N(T) = \frac{N_{\rm A}\mu_0g^2\mu_{\rm B}^2}{Nk_{\rm B}T}
    \left(\left\langle\left(S_z^{\rm tot}\right)^2\right\rangle
    - \left\langle S_z^{\rm tot}\right\rangle^2\right),
\end{equation}
where $\langle\dots\rangle$ denotes the thermal average, $N_{\rm A}$
is the Avogadro constant, $\mu_0$ the magnetic permeability, $g$ the
gyromagnetic ratio, $\mu_{\rm B}$ the Bohr magneton, and $k_{\rm B}$
the Boltzmann constant.  For the present nonmagnetic system
$\left\langle S_z^{\rm tot}\right\rangle=0$.

\begin{figure}
    \includegraphics[width=.8\columnwidth]{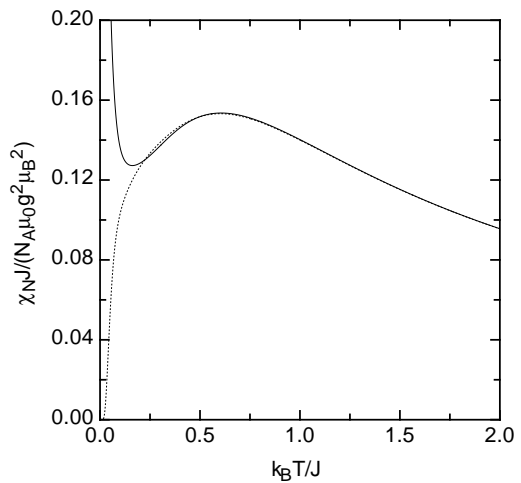}
    \caption{Temperature dependence of the uniform magnetic
    susceptibility of Heisenberg chains of length 23 (solid line) and
    24 (dotted line) with open boundary conditions.}
    \label{fig:fig4}
\end{figure}
To illustrate the results, Figure~\ref{fig:fig4} shows the
temperature dependence of the magnetic susceptibility for chains of
length 23 and 24.  According to Bonner and Fisher~\cite{bonner:64}, the
maximum of the susceptibility for an infinite Heisenberg chain is
given by $\chi_\infty^{\rm max}J/(N_{\rm A}\mu_0g^2\mu_{\rm
B}^2)\approx0.147$ and is reached at a temperature of $k_{\rm B}T_{\rm
max}/J\approx0.641$.  For the 24-site chain, the corresponding results
are 0.153 for the maximum susceptibility, and 0.601 for the position
of the maximum.

\begin{figure}
    \includegraphics[width=.8\columnwidth]{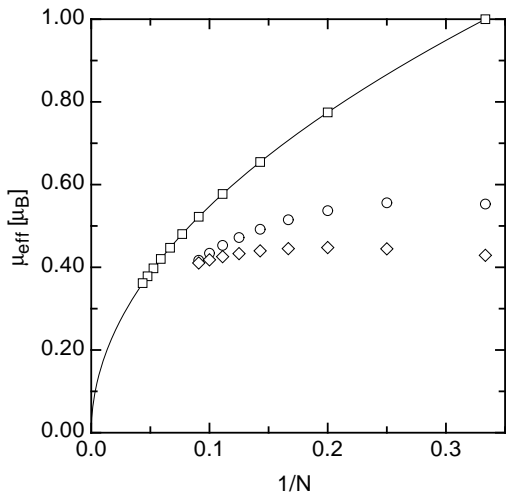}
    \caption{Calculated effective moments in units of the Bohr
    magneton as a function of the inverse chain length.  Squares
    denote the results for chains of a single fixed length, circles
    for the Wigner average, diamonds for the Poisson average with mean
    $N$ over chain lengths between two and 24.  The line denotes the
    dependence $\mu_{\rm eff}(N)=\sqrt{3/N}\,\mu_{\rm B}$.}
    \label{fig:fig5}
\end{figure}
From the low-temperature upturn of $\chi_N$ ($N$ odd), we have
extracted the effective moment $\mu_{\rm eff}(N)$ for each value of
$N$.  These data are shown in Figure~\ref{fig:fig5} as the open
squares.  The effective moments as a function of the inverse chain
length scale perfectly according to $\mu_{\rm
eff}(N)=\sqrt{3/N}\,\mu_{\rm B}$.  The factor $\sqrt{3}$ originates
from the definition of the effective moment via Curie's law,
$\chi_{\rm C}=N_{\rm A}\mu_0\mu_{\rm eff}^2/(3k_{\rm B}T)$.

The effective moment as a function of the mean chain length $N$ after
averaging over chains of different size (even and odd) is shown in
Figure~\ref{fig:fig4} as open circles, representing the Wigner
distribution and diamonds, representing the Poisson distribution. 
Data are shown for $N\le11$, because for a larger mean length, the
chains of size greater than 24 would contribute significantly to the
average susceptibility.

The main effect for both averages is to reduce the magnetic moment for
short mean chain lengths.  For the Wigner distribution, this is due to
the linear suppression of the weight of the susceptibilities for the
shortest chains, whereas for the Poisson distribution, the moment
reduction is due to a comparatively large weight of the longer chains
in the partition sum.  Independent of how the average is performed,
the effective moment is reduced because chains with an even site
number, and thus no Curie term at low $T$, are also included in the
average.  The data suggest that for mean chain lengths $N$ larger than
10, the average effective moment depends only weakly on the details of
defect distribution.

\begin{figure}
    \includegraphics[width=.8\columnwidth]{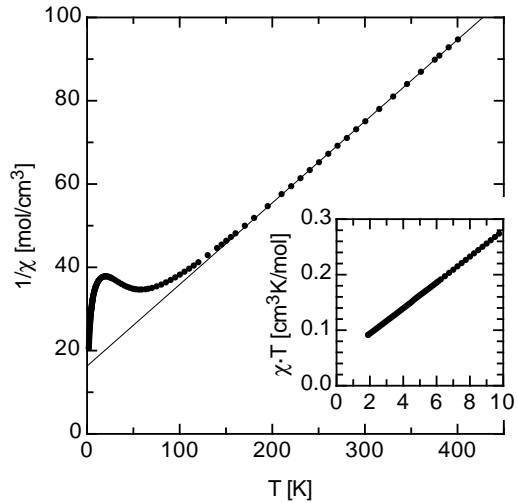}
    \caption{Curie plot of the susceptibility of Ba$_2$V$_3$O$_9$. 
    The straight line illustrates a fit of the form
    $\chi=C_0/(T+\Theta_{\rm CW})$ for temperatures $T$ between 200\,K
    and 400\,K. In the inset, the same data are plotted as
    susceptibility times temperature versus $T$ for low temperatures.}
    \label{fig:fig2}
\end{figure}
In Figure~\ref{fig:fig2}, the experimental data for the susceptibility
$\chi$ of Ba$_2$V$_3$O$_9$ are plotted as $\chi\cdot T$ versus
temperature $T$.  A fit to these between $T=2\,\rm K$ and 10\,K of the
form $\chi=C/T+\chi_{\rm VV}$ with $C=N_{\rm A}\mu_0\mu_{\rm
eff}^2/(3k_{\rm B})$ yields an average moment of $\mu_{\rm
eff}\approx0.2\mu_{\rm B}$.  A mean chain length of the order of 75
spins would be necessary to explain the experimentally observed Curie
upturn by the segmentation of VO$_2$ chains.

\section{Conclusion}
We have discussed two alternative models for the anomalous Curie term
in the low-$T$ susceptibility of Ba$_2$V$_3$O$_9$.  We find that
within our approximate calculation the staggered field mechanism is
far too small to explain the observed effective moment.  Therefore we
propose a different mechanism for the Curie term which is based on
chain segmentation due to broken magnetic bonds which leads to
uncompensated effective moments on segments with odd number of spins. 
We find that a rather low concentration of $\sim 10^{-2}$ broken bonds
can explain the observed Curie term.  A definite distinction between
the two models requires the investigation of the field-orientation
dependence of the susceptibility in monocrystals of Ba$_2$V$_3$O$_9$.

\acknowledgments

We wish to thank Henk Eskes for supplying his exact-diagonalization
routines which were included in the finite-temperature Lanczos code
used here.  We thank Enrique Kaul for discussions and for supplying
his experimental data.

\bibliography{references}

\end{document}